\documentclass[12pt]{article}
\usepackage{color}
\usepackage{epsfig}
\usepackage{amsmath}
\usepackage{hhline}
\usepackage{amsfonts,amssymb}
\usepackage{times}

\graphicspath{{./eps/}}
\DeclareGraphicsExtensions{.eps,.eps.gz,.ps,.ps.gz}
\newlength{\dinwidth}
\newlength{\dinmargin}
\newcommand{\cO}{{\cal O}}

\newcommand{\hdick}{\noalign{\hrule height1.4pt}}

\newcommand{\GeV} {\mathrm{GeV}}

\newcommand{\fb}  {\mathrm{fb}}
\newcommand{\fbi} {\mathrm{fb}^{-1}}

\newcommand{\mrad} {\mathrm{mrad}}

\newcommand{\cL } {{\cal L}}
\newcommand{\cP } {{\cal P}}

\def\susy    {{\sc Susy}}

\def\pythia  {{\sc Pythia}}

\def\tesla  {{\sc Tesla}}

\def\ee{e^+e^-}
\def\ti    {\tilde}

\def\stau  {{\ti\tau}}

\def\sell  {{\ti\ell}}

\def\cx    {\ti {\chi}}

\def\nt    {\ti {\chi}^0}

\def\smur  {{\ti\mu}_R}

\def\smurm {{\ti\mu}^-_R}

\def\smurp {{\ti\mu}^+_R}

\def\sel   {{\ti e}_L}
\def\ser   {{\ti e}_R}

\def\serm  {{\ti e}^-_R}

\def\serp  {{\ti e}^+_R}

\def \Eslash {E \kern-.75em\slash }
\def \Mslash {M \kern-.5em\slash }

\newcommand{\beq}{\begin{equation}}
\newcommand{\eeq}{\end{equation}}
\newcommand{\bea}{\begin{eqnarray}}
\newcommand{\eea}{\end{eqnarray}}

\newcommand{\fig}[1]{figure~\ref{#1}}
\newcommand{\tab}[1]{table~\ref{#1}}

\begin{document}

\title{ 
  \begin{flushright} \normalsize
      August 2004 \\[1em]
  \end{flushright} 
  \LARGE\bfseries
  Detection of sleptons at a linear collider
  in  models with small slepton-neutralino
  mass differences\footnote{
    Contribution to the ECFA Study on Physics and
    Detectors for a Linear Collider and 
    the International Conference on Linear Colliders,
    LCWS~04, Paris, April 2004}
    }
\author{\Large Hans-Ulrich Martyn\footnote{email: martyn@mail.desy.de}
  \\[.5ex]  
  { \itshape 
     I. Physikalisches Institut, RWTH Aachen, Germany}}   
\date { }
 
\maketitle
\thispagestyle{empty}

\begin{quote}   \small
A feasibility study is presented to precisely measure the masses of 
charged sleptons and the lightest neutralino
in $\ee\to\smur\smur$ and $\ee\to\stau_1\stau_1$ production.
Different mSUGRA scenarios with small slepton-neutralino mass
differences of a few GeV are invesigated.
The analysis is based on a detailed simulation of the final state
lepton energy spectra including background reactions and assuming realistic
experimental conditions at the \tesla \ $\ee$ Linear Collider.
Effects on the mass resolutions due to beams colliding head-on or
under a crossing angle are discussed.
\end{quote}

\section{Introduction}

In the minimal supersymmetric standard model the lightest
superparticle (LSP), which is stable in $R$--parity conserving scenarios, is
a natural candidate for cold dark matter. 
Measurements of the {\sc Wmap}\
experiment~\cite{wmap} have determined the amount of cold dark matter in
the universe to be $\Omega_{CDM} h^2 = 0.113\pm0.009$.
As a consequence of this result, considerable constraints on the
parameters of supersymmetric theories can be set~\cite{baer}. 
For example, in minimal supergravity (mSUGRA), where usually the
LSP is the neutralino $\nt_1$,
the favoured particle spectra are typically characterised 
by small mass differences between the
lightest slepton (usually the stau)
and neutralino (coannihilation region) or
between the light chargino and neutralino (focus point region). 
Collider experiments will be essential to test the supersymmetry dark
matter hypothesis. But only a future $\ee$ Linear Collider will be
able to provide precise enough mass measurements of the light
\susy\ particles
in order to compute the relic density of the LSP to an accuracy
comparable with {\sc Wmap}\ and other proposed experiments.

Experimentally challenging is the coannihilation region where the
slepton--neutralino mass difference $\Delta m$
may be as small as a few GeV.
In this note  we investigate the capabilities of a TeV $\ee$ Linear
Collider like \tesla~\cite{tdr} to detect and measure $\smur\smur$ and
$\stau_1\stau_1$ pair production in models with small  $\Delta m$
in order to  determine  precisely
the masses of the light sleptons as well as of the neutralino $\nt_1$
from the energy spectra of the final states.
The case studies are based on spectra from a SPS~1a inspired
model~\cite{spsmodels}, 
where the neutralino mass has been increased 
to accommodate a small mass difference $\Delta m$
and from the benchmark model D' proposed in \cite{pwmap}.

The most severe background comes from four fermion production via
two-photon ($\gamma\gamma$) processes,
$\ee\to\ee f \bar f$, 
and it is important to have an
efficient veto of the spectator electrons/positrons down to very small
scattering angles.
The present study also addresses the consequences for background
suppression 
in head-on collisions, as envisaged in the \tesla \ design, 
versus the option of
having a crossing angle of $2\cdot 10~\mrad$ between the colliding
beams.

\section{Event generation}
 
The processes under study are the pair production of scalar muons and 
scalar taus 
\begin{eqnarray}
  e^+_L e^-_R & \to & \smur\smur 
              \ \to \ \mu^+\nt_1\,\mu^-\nt_1 ,
  \label{smurproduction}  \\   
  e^+_L e^-_R & \to & \stau_1\stau_1 
              \ \to \ \ \ \tau^+\nt_1\,\tau^-\nt_1 \ .
  \label{stauproduction}
\end{eqnarray}
The beams are assumed to be longitudinally polarised, right-handed
electrons $e^-_R$ and left-handed positrons $e^+_L$,
which considerably increase the production cross sections and suppress
background. 

The slepton decay $\sell^- \to \ell^- \nt_1$
into an ordinary lepton and neutralino
is isotropic and produces a flat lepton energy spectrum.
The endpoint energies $ E_{+/-}$ are related to the primary
slepton and the neutralino masses 
(neglectging the lepton mass) 
\begin{eqnarray}  
  E_{+/-} & = &
        \frac{\sqrt{s}}{4} 
        \left ( \frac{m_{\sell}^2 - m_{\cx}^2}{m_{\sell}^2} \right ) 
        \, \left (1 \pm \sqrt{1-4\,m_{\sell}^2/s} \: \right ) \ ,  
	\label{eq_endpoints} \\[.5ex]
        m_{\sell} & = & \sqrt{s} \,
        \frac{\sqrt{E_{-}\,  E_{+} }}{E_{-}+E_{+}} \ ,
	\label{eq_mslepton} \\[.5ex]
        m_{\cx} & = & m_{\sell} \,
          \sqrt{1 - \frac{E_{-}+E_{+}}{\sqrt{s}/2}} \ .
	  \label{eq_mneutralino}
\end{eqnarray}
One observes that with decreasing slepton-neutralino mass difference
$\Delta m$ the endpoint energies get lower and come closer together.
For $\tau$ decays the spectrum is quite distorted,
the maximal energy of the observable particles
coincides with the upper edge $E_+$, but the lower edge at $E_-$ 
is completely diluted. 
However, the shape of the energy distribution can still be used to
extract the $\stau$ mass with high precision.

Events are generated with the program \pythia~6.2~\cite{pythia}
which includes beam polarisations $\cP_{e^\pm}$, 
initial and final state QED radiation as well as beamstrahlung \`a la
{\sc Circe}~\cite{circe}.
The decays of $\tau$ leptons are treated by {\sc Tauola}~\cite{tauola}.

\begin{figure}
\begin{picture}(100,70)(0,0) \footnotesize
  \put(0,0){
    \put(0,0){\epsfig{file=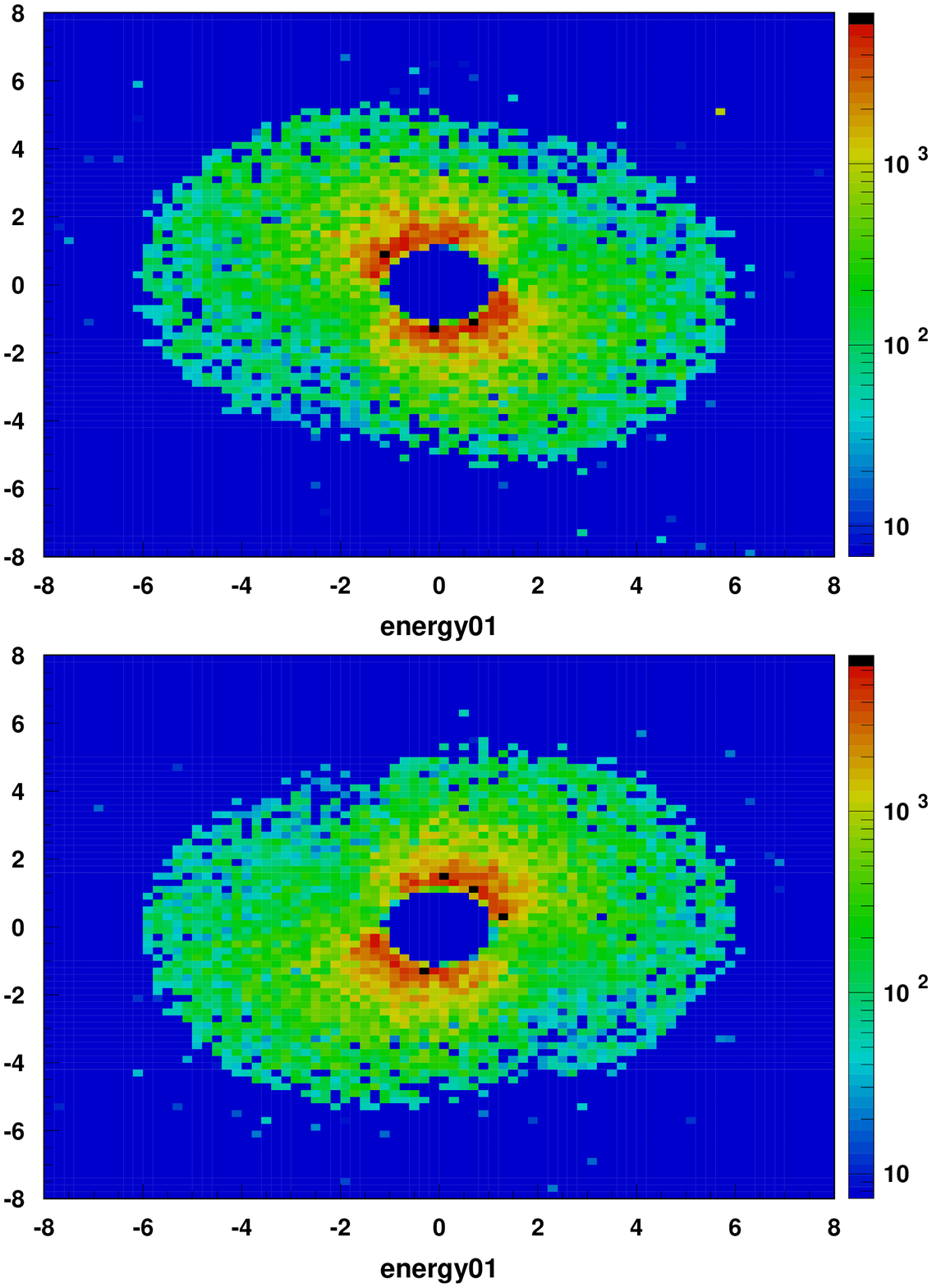,%
	  bbllx=10pt,bblly=400pt,bburx=590pt,bbury=750pt,clip=,%
	  width=.45\textwidth,height=.4\textwidth} }
      \put(0,-72){\epsfig{file=box.eps,
	  width=.45\textwidth,height=.6\textwidth}}
      }
  \put(70,0){
     \put(0,0){\epsfig{file=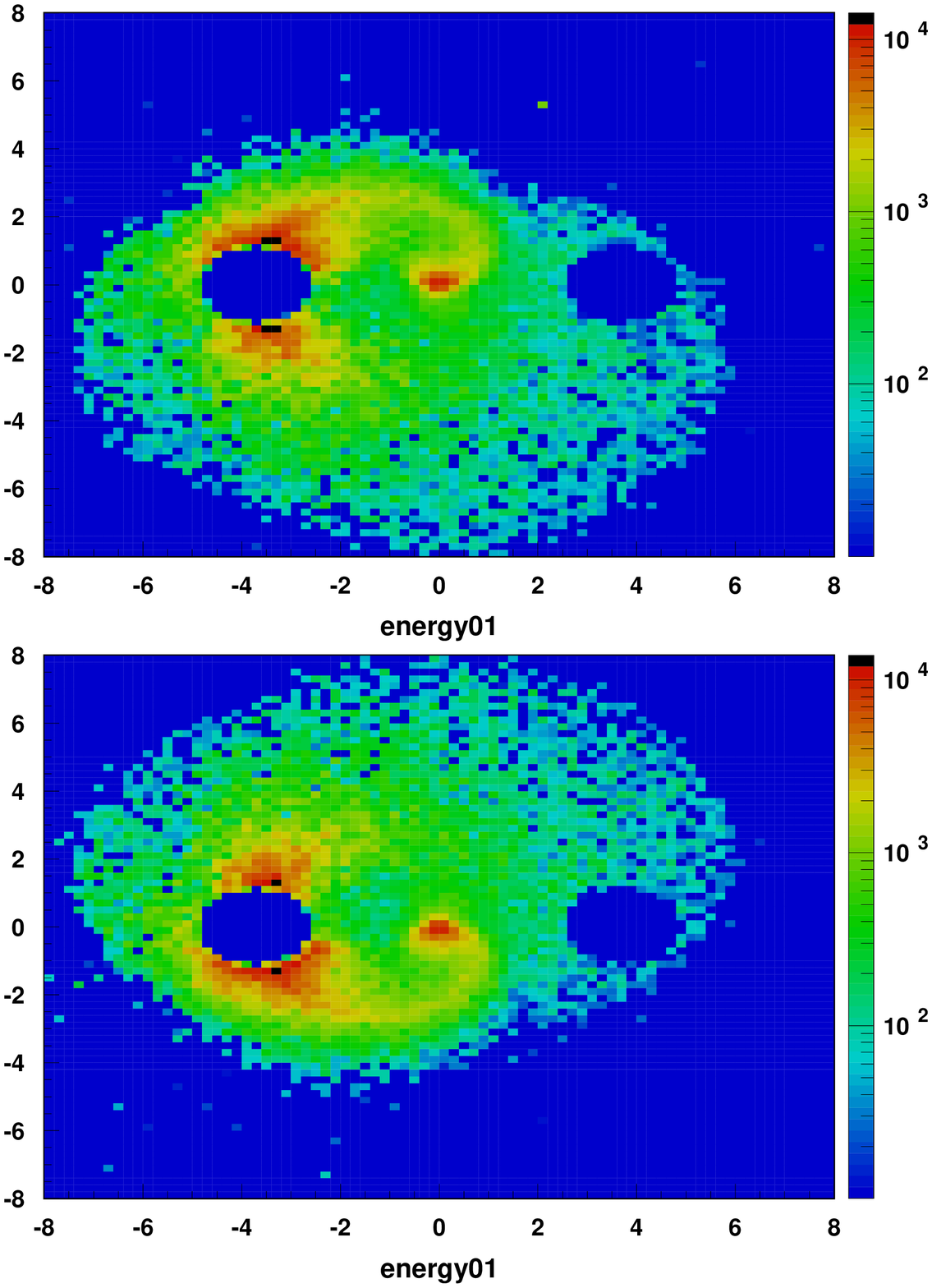,%
	  bbllx=10pt,bblly=400pt,bburx=590pt,bbury=750pt,clip=,%
	  width=.45\textwidth,height=.4\textwidth} }
      \put(0,-72){\epsfig{file=box.eps,
	  width=.45\textwidth,height=.6\textwidth}}
     }
  \put(10,65){\bf head-on collision}
  \put(80,65){\bf crossing $\delta\theta_c = 2\cdot 10$~mrad}
  \put(58,63){\scriptsize ${\mathrm{E~[GeV/cm}}^{2}]$}
  \put(128,63){\scriptsize ${\mathrm{E~ [GeV/cm}}^{2}]$}
  \put(120,0){\scriptsize  ${\mathrm{x~ [cm]}  } $}
  \put( 50,0){\scriptsize  ${\mathrm{x~ [cm]}  } $}
  \put( 0,57){\scriptsize  ${\mathrm{y~ [cm]}  } $}
  \put(70,57){\scriptsize  ${\mathrm{y~ [cm]}  } $}
\end{picture}
\label{bunchxing}
\caption{Energy deposits $[\GeV/{\rm cm}^2]$
  from beamstrahlung at $\sqrt{s}=500~\GeV$
  for head-on collisions (left) and crossed beams (right) 
  in a plane perpendicular to the beams at  
  3.7~m from the interaction point~\cite{buesser} }
\end{figure}

The  detector simulation is based on the detector proposed
in the {\sc Tesla tdr}~\cite{tdr}
and implemented in the Monte Carlo program {\sc Simdet}~4.02~\cite{simdet}.
The main detector features are excellent particle identification and
measurement 
for a polar angle acceptance $\theta > 125~\mrad$. 
The forward region\footnote{The acceptance at an angle $\theta$
  in the forward region is symmetric in the backward direction at
  $\pi-\theta$}  
is equipped with electromagnetic calorimeters in
order to effectively veto electrons and photons down to very small
scattering angles. 
For $\theta < 27.5~\mrad$ a highly segmented LCAL calorimeter
situated at a distance of 3.7~m from the interaction point has
recently been proposed~\cite{buesser}. 
For head-on collisions, as envisaged at \tesla,
a beam pipe of radius $r=1.2$~cm appears feasible and
the calorimetric coverage may start at $3.5~\mrad$ 
(assuming 1~mm clearance).
For beams colliding under a crossing angle of
$2\cdot 10~\mrad$ 
there are two dead regions and the exiting, disrupted beams
require a larger beampipe of $r=2.0$~cm,
giving a minimal tagging angle of $5.7~\mrad$. 
In both cases $e, \ \gamma$ tagging at low angles is very demanding above 
the huge background of energy deposits caused by beamstrahlung, as 
shown in \fig{bunchxing}~\cite{buesser}. 
The beamstrahlung halo varies rapidly around the beam pipes.
Further, the angle between the magnetic field and the crossed beams
leads to an enhanced (factor 2) and asymmetric energy distribution.
An algorithm based on topological properties of electromagnetic
showers is used to veto electrons and photons a few mm close to the
beam pipe for energies above $50~\GeV$ with high efficiency~\cite{drugakow} 
in both designs. 

For \tesla\ a geometry with crossed beams is an option which may
facilitate beam diagnostics. A normal conducting linear collider like
{\sc Nlc/Glc} has to be operated with a crossing angle and the short
bunch separation may lead to additional, disturbing pile up effects,
depending on the calorimeter read-out. 
This difficulty will not be further discussed.

\section{Case study SPS~1a inspired model}

The aim of this paper is to investigate the experimental problems
related to small lepton-neutralino mass differences and to compare the
results with more conventional benchmark spectra.
As reference serves the  mSUGRA scenario SPS~1a where detailed
studies of slepton production in the continuum~\cite{martyn} and at
threshold~\cite{freitas} exist.
The case study is based on a SPS~1a inspired model, where the masses of
the the light sleptons are kept unchanged,
$m_{\smur}=143~\GeV$ and $m_{\stau_1}=133~\GeV$,
and the neutralino masses are increased. 
Within mSUGRA a $\stau_1 - \nt_1$
mass splitting $\Delta m_{\stau} = 8~\GeV$
can {\it e.g.} be accomplished by shifting the common
gaugino mass $m_0$ to $70~\GeV$ and the common scalar mass $m_{1/2}$
to $320~\GeV$ 
and leaving the other parameters as $\tan\beta=10$, $A_0 = -100~\GeV$
and sign~$\mu>0$.

The strategy will be to determine the neutralino mass from the
lepton spectra of $\smur\smur$ and/or $\ser\ser$ production
(eqs.~(\ref{eq_mneutralino}) and (\ref{eq_mslepton}))
and then
use this value as input in the $\tau$ decay spectra of the $\stau$
analysis. 
Optimal collider operating conditions are chosen such as to maximise
the signal cross section and to minimise the background:
a center of mass energy of $\sqrt{s} = 400~\GeV$ and
beam polarisations of $\cP_{e^-} = +0.8$ for right-handed electrons 
and of $\cP_{e^+} = -0.6$ for left-handed positrons.
Other \susy \ processes are kinematically not accessible 
(below production thresholds)
or strongly suppressed ($\stau_1\stau_2$, $\ser\sel$ and $\nt_1\nt_2$).
The SM contribution from $WW$ production is negligible due to the beam
polarisation and efficient selection criteria. 
The essential remaining background comes from two-photon processes
$e^+e^- \to e^+e^-  \ell^+\ell^-$ with cross sections of
$\cO(10^6~\fb)$.
The simulations assume an integrated luminosity of $\cL = 200~\fbi$,
corresponding to about one year of data taking.

\subsection{Detection of {\boldmath $e^+ e^- \to\smur\smur$}}

For the purpose of illustration a small smuon-neutralino mass
difference of $\Delta m=8~\GeV$ is assumed.
Such a situation may occur for low values of $\tan\beta$ with the light
smuon and stau being almost degenerate.
The masses used in the simulation of 
$e^+_L e^-_R\to \smurp \smurm \to \mu^+ \nt_1 \,\, \mu^-\nt_1 $
are $m_{\smur}=143.0~\GeV$ and $m_{\nt_1}=135.0~\GeV$. 
The relevant signal and background cross sections are 
$\sigma_{\smur\smur} = 120~\fb$,
$\sigma_{WW}=1000~\fb$ and $\sigma_{ee\mu\mu}=9.4\cdot10^5~\fb$.

The following event selection criteria are applied:
(1) veto of any electromagnetic energy ($e$ or $\gamma$) above 5~GeV in
    the forward region $\theta<125~\mrad$,
(2) two muons within the polar angle acceptance
    $-0.90 < Q_\mu \cos\theta_\mu < 0.75$,
(3) acoplanarity angle $\Delta\phi^{\mu\mu} < 160^\circ$,
(4) missing momentum vector inside active detector
    $\cos\theta_{\vec p_{miss}} < 0.9$,
(5) muon energy $ E_\mu > 2\;\GeV$,
(6) transverse momentum of di-muon system $p_\perp^{\mu\mu} > 5\;\GeV$. 
The overall signal efficiency is $60\;\%$, while the remaining
background is negligible.
The most effective cuts against $WW$ production are (2) and (3).
The $\gamma \gamma\to\mu\mu$ background is reduced by a factor of
$\sim 10^{-6}$, particularly powerful are cuts (3) and (6).

\begin{figure}[htb]
\begin{picture}(150,90)  (0,10) 
  \put(0,0){\epsfig{file=eptcut.smu.eps,
      bbllx=250pt,bblly=220pt,bburx=550pt,bbury=440pt,clip=,%
      width=.65\textwidth,angle=90} }
  \put(0,0){\epsfig{file=eptcut.eemmh.eps,
       bbllx=250pt,bblly=220pt,bburx=550pt,bbury=440pt,clip=,%
     width=.65\textwidth,angle=90} }

  \put(80,0){\epsfig{file=eptcut.smu.eps,%
      bbllx=250pt,bblly=220pt,bburx=550pt,bbury=440pt,clip=,%
      width=.65\textwidth,angle=90} }
  \put(80,0){\epsfig{file=eptcut.eemm.eps,
      bbllx=250pt,bblly=220pt,bburx=550pt,bbury=440pt,clip=,%
      width=.65\textwidth,angle=90} }
  \put( 17,97){\epsfig{file=box.eps,width=85pt,height=20pt}}
  \put( 97,97){\epsfig{file=box.eps,width=85pt,height=20pt}}

  \put( 20,91){\color{blue} \bf head-on collision }
  \put(100,91){\color{green}\bf crossed beams }
\end{picture}
  \caption{Muon energy spectra $E_\mu$ from the reactions 
    $e^+_L e^-_R \to \smurp\smurm \to \mu^+\nt_1 \; \mu^-\nt_1$
    (points)
    and 
    $e^+_L e^-_R\to e^+ e^- \,\mu^+\mu^-$ (histogram)
    without (left) and with (right) beam crossing angle.
    SPS~1a inspired model, $m_{\smur} = 143\;\GeV, \ \Delta m = 8\;\GeV$,
    $\sqrt{s}=400\;\GeV$ and $\cL=200\;\fbi$}
  \label{smur_spectrum}
\end{figure}

The resulting spectra of the muon energy $E_\mu$ 
are shown in \fig{smur_spectrum}. 
The rectangular shape with the steeply rising edges is clearly observable. 
The background is roughly a factor of 2 larger in a configuration
with beam crossing angle compared with head-on collisions.
However, the influence on the endpoint energy
measurements are in both cases negligible.
A fit to the spectrum yields the endpoint energies
$E_- = 3.270 \pm 0.015\;\GeV$ and $E_+ = 18.480 \pm 0.39\;\GeV$,
from which the strongly correlated masses of $\smur$ and $\nt_1$ are
derived with an accuracy of
$\delta m_{\smur} = 0.18~\GeV$ and $\delta m_{\nt_1} = 0.17~\GeV$,
see \tab{tab_sleptons}.

The precision on the neutralino mass can be considerably improved by
analysing the selectron production 
$e^+_L e^-_R \to \serp\serm \to e^+\nt_1\,e^-\nt_1$. 
The cross section is roughly a factor of four higher while the two-photon
background remains low.  
From the results of \tab{tab_sleptons} one concludes that a neutralino
mass resolution of $\delta m_{\nt_1} = 0.08~\GeV$ is achievable, which
can be further improved by a combined analysis of both slepton
production processes.

\begin{table}[htb] \centering \vspace*{2mm}
  \fbox{
    \begin{tabular}{l c c}
      $\ee \to\smur\smur$ & $ m_{\smur} = 143.0 \pm 0.18~\GeV$
                     & $ m_{\nt_1} = 135.0 \pm 0.17~\GeV$ \\
      $\ee \to\ser\ser$   & $m_{\ser} = 143.0 \pm 0.09~\GeV$
	             & $ m_{\nt_1} = 135.0 \pm 0.08~\GeV$
    \end{tabular} }    
  \caption{Expected mass resolutions for $\smur$, $\ser$ and $\nt_1$ 
    from lepton energy spectra in a SPS~1a inspired scenario 
    assuming $\cL=200\;\fbi$ at $\sqrt{s}=400~\GeV$}
  \label{tab_sleptons}
\end{table}

Compared with the original SPS~1a scenario~\cite{martyn},
the mass resolutions obtained from the first and second generation
slepton pair production do not degrade with decreasing slepton-neutralino
mass difference down to values of $\Delta m = 8~\GeV$, and possibly
even lower to $5~\GeV$.

\subsection{Detection of {\boldmath $e^+ e^-\to \stau_1 \stau_1$} }

For the simulation of 
$e^+_L e^-_R\to \stau_1^+ \stau_1^- \to\tau^+\nt_1 \,\,\tau^-\nt_1$
a stau-neutralino mass difference of $\Delta m=8~\GeV$ is assumed,
leading to  $m_{\stau_1}=133.2~\GeV$ and $m_{\nt_1}=125.2~\GeV$.
The cross section is $\sigma_{\stau_1\stau_1} = 140~\fb$,
SM background considered have cross sections of
$\sigma_{WW}=1000~\fb$ and $\sigma_{ee\tau\tau}=4.5\cdot10^5~\fb$.
For the identification and reconstruction
of $\tau$ pairs, one tau is required to decay
into hadrons while the other may decay hadronically or leptonically.
The measurement of energy spectra is restricted to the hadronic 
decay modes 
$\tau\to\pi\nu_\tau\; (11.1\%)$,  
$\tau\to\rho\nu_\tau \to\pi^\pm\pi^0\nu_\tau \;(25.4\%)$ and
$\tau\to 3\pi\nu_\tau \to \pi^\pm \pi^+\pi^-\nu_\tau + 
  \pi^\pm\pi^0\pi^0 \nu_\tau \;(19.4\%)$.
A difficulty may pose the $\tau$ polarisation from 
$\stau_1\to\tau\nt_1$ decays, which has an influence
on the pion energy spectrum, but does not affect the shape
of the $\rho$ and $3\pi$ spectra~\cite{nojiri}.
The $\tau$ polarisation can either be taken from calculations
or can be determined simultaneously~\cite{martyn},
if one wants to keep the pion decay channel.
The leptonic 3-body decays 
$\tau\to e\nu_e\nu_\tau$, $\tau\to\mu\nu_\mu\nu_\tau$
exhibit a very weak dependence on $m_\stau$ and are therefore
discarded in the analysis.

\begin{figure}[htb]
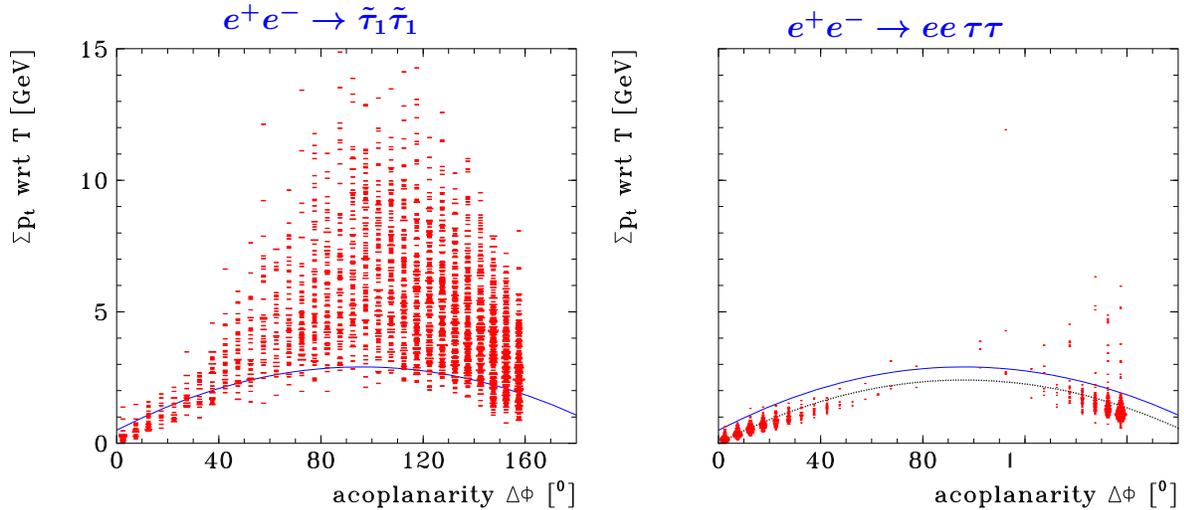

  \begin{picture}(150,75)  \boldmath
    \put(0,5) {\epsfig{file=sumpt_stau.eps,%
	bbllx=0pt,bblly=0pt,bburx=530pt,bbury=680pt,clip=,%
	width=.4\textwidth,angle=90} }
    \put(80,5){\epsfig{file=sumpt_eett.eps,%
	bbllx=0pt,bblly=0pt,bburx=530pt,bbury=680pt,clip=,%
	width=.4\textwidth,angle=90} }
    \put( 30,69){{\color{blue} $\ee \to \stau_1 \stau_1 $}}
    \put(105,68){{\color{blue} $\ee \to ee\, \tau \tau $}}
  \end{picture}
  \caption{Correlations of 
    $\sum p^{\tau}_{\perp,\ \vec T}$ wrt thrust axis $\vec T_\perp$
    versus acoplanarity angle
    $\Delta\phi^{\tau\tau}$ 
    in a plane perpendicular to the beams for 
    $e^+_L e^-_R\to \stau_1  \, \stau_1   
    \to \tau^+ \nt_1 \,\, \tau^-\nt_1 $ (left) and
    $e^+_L e^-_R\to e^+e^- \tau^+  \, \tau^-$ (right). 
    The curves indicate the separation between \susy\ and
    $\gamma\gamma$  processes.
    SPS~1a inspired model, $m_{\stau_1} = 133\;\GeV, \ \Delta m = 8\;\GeV$,
    $\sqrt{s}=400\;\GeV$} 
  \label{sumpt_spectrum}
\end{figure}

The following criteria are applied to select $\stau_1\stau_1$ events:
(1) veto of any electromagnetic energy ($e$ or $\gamma$) above 5~GeV in
    the forward region $\theta<125~\mrad$,
(2) two $\tau$ candidates within the polar angle 
    $-0.75 < \cos\theta_\tau < 0.75$,
(3) invariant mass of $\tau$ jet $ m_\tau < 2\;\GeV$,
(4) tau energy $ 2 < E_\tau < 25\;\GeV$,
(5) acoplanarity angle $\Delta\phi^{\tau\tau} < 160^\circ$,
(6) missing momentum vector 
    $\cos\theta_{\vec p_{miss}} < 0.8$,
(7) transverse momentum of di-tau system 
    $3 < p_\perp^{\tau\tau} < 25\;\GeV$,
(8) combined cut on $\sum p_{\perp,\ \vec T}^\tau$ 
    and $\Delta\phi^{\tau\tau}$.
An effective reduction of $\gamma\gamma$ background is possible by
using observables defined in the plane perpendicular to the beams.
Cuts (5) and (7) are not as efficient as in the smuon analysis.
They have to be supplemented by criterion (8),  illustrated in
\fig{sumpt_spectrum} as correlations between 
 $\sum p_{\perp,\ \vec T}^\tau$, 
the sum of $\tau$ momenta projected onto the transverse thrust axis 
$\vec T_\perp$, and the acoplanarity angle  $\Delta\phi^{\tau\tau}$.
A cut as indicated by the curves has little influence on the signal
and reduces the background by factors of order 50.
The residual $\stau_1\stau_1$ efficiency is $\sim 18\%$. 
The two-photon background is suppressed by 
$\sim 1.5\cdot 10^{-6}$ for head-on collisions,
respectively $\sim 2.5\cdot 10^{-6}$ in the case of crossed beams.
Contributions from $WW$ production are negligible.

The energy spectra of  $E_\pi$,  $E_\rho$ and  $E_{3\pi}$ are shown in
\fig{etau_spectrum} for collisions with crossed beams.
Clean signals above a very small and controllable background are observed.
The spectra are fitted to full simulations with different $\stau_1$
mass hypothesis, yielding a sensitivity of
$\delta m_{\stau_1} = 0.14~\GeV \oplus \delta m_{\nt_1}$ 
for the $E_\pi$ spectrum, and
$ \delta m_{\stau_1} = 0.10~\GeV \oplus \delta m_{\nt_1}$ for the
 $E_\rho$ and $E_{3\pi}$ spectra, each.
Assuming a conservative estimate of $\delta m_{\nt_1} = 0.10~\GeV$ one
expects a combined uncertainty of 
$ \delta m_{\stau_1} = 0.14~\GeV$ for the the stau mass.
A lower background expected in head-on collisions 
or a twofold higher rate (accounting for additional two-photon
processes not covered) do not change the resolution significantly.

\begin{figure}[htb]
\begin{picture}(150,80)  (0,0)
  \put(-10,-70){
    \put(0,0){\epsfig{file=ehadx.stau.eps,width=.925\textwidth,angle=90} }
    \put(17,138){\color{blue} \boldmath$\tau\to\pi\nu$}
    \put(76,138){\color{blue} \boldmath$\tau\to\rho\nu$}
    \put(135,138){\color{blue}\boldmath$\tau\to 3\pi\nu$}
  }     
\end{picture}
  \caption{Hadron energy spectra $E_\pi$ of $\tau\to\pi\nu_\tau$,
    $E_\rho$ of $\tau\to\rho\nu_\tau$ and
    $E_{3\pi}$ of $\tau\to 3\pi\nu_\tau$ decays
    from the reactions
    $e^+_L e^-_R\to \stau_1  \, \stau_1   
    \to \tau^+ \nt_1 \,\, \tau^-\nt_1 $ and 
    $e^+_L e^-_R\to e^+e^- \tau^+\tau^-$
    assuming crossed beams collision.
    SPS~1a inspired model, $m_{\stau_1} = 133.2\;\GeV, \ \Delta m = 8\;\GeV$,
    $\sqrt{s}=400\;\GeV$ and $\cL=200\;\fbi$}
  \label{etau_spectrum}
\end{figure}

It is interesting to note that the precision achieved in the present
study is about twice as good as for the original SPS~1a
scenario~\cite{martyn}. The reason is that for small stau-neutralino
mass differences the energy spectra are narrower and are falling
steeper at the edges.
In particular, a determination of the high endpoint energy,
eq.~(\ref{eq_endpoints}), is therefore more precise, which directly
enters the mass calculation.
      
\begin{figure}[htb]
\begin{picture}(150,80) 
  \put(-10,-70){
    \put(30,70){\epsfig{file=ehadhon.staudm5.eps,%
	bbllx=265pt,bblly=210pt,bburx=570pt,bbury=430pt,clip=,%
	width=.5\textwidth,angle=90} }
    \put(90,70){\epsfig{file=ehadhon.staudm3.eps,
	bbllx=265pt,bblly=210pt,bburx=570pt,bbury=430pt,clip=,%
	width=.5\textwidth,angle=90} }
    \put(115,145){\color{black}\footnotesize$\Delta m = 3\,\GeV$}
    \put( 55,145){\color{black}\footnotesize$\Delta m = 5\,\GeV$}
    \put( 43,139){\color{blue}\boldmath$\tau\to\rho\nu$}
    \put(103,139){\color{blue}\boldmath$\tau\to\rho\nu$}
  }     
\end{picture}
  \caption{Hadron energy spectra 
    $E_\rho$ of $\tau\to\rho\nu_\tau$  from 
    $e^+_L e^-_R\to \stau_1  \, \stau_1   
    \to \tau^+ \nt_1 \,\, \tau^-\nt_1 $ 
    with $\Delta m = 5\,\GeV$ (left) and $\Delta m = 3\,\GeV$ (right) 
    assuming head-on collision.
    SPS~1a inspired model, $m_{\stau_1} = 133.2\;\GeV$, 
    $\sqrt{s}=400\;\GeV$ and $\cL=200\;\fbi$}
  \label{erhodm_spectrum}
\end{figure}

Examples of a further reduction of the stau-neutralino mass difference
to $\Delta m = 5~\GeV$ and  $\Delta m = 3~\GeV$ are shown for the
$\rho$ energy spectra in \fig{erhodm_spectrum}. 
The overall event rates decrease but are still sufficient to clearly
determine the high endpoint energy. Including mass effects, the
maximal $\tau$ energy and its relation to the $\stau_1$ mass are given by
\begin{eqnarray} 
    E_+ & = & \frac{m^2_{\stau_1} - m^2_{\nt_1}}{2\,(E_{\stau_1}
    -p_{\stau_1}f_{\tau})}   
    \qquad {\rm with } \quad f_{\tau} = 1-0.5\,(m_\tau/\Delta m)^2 \ ,
    \\[.4ex]
    \delta m_{\stau_1} & \simeq & 0.38\, \delta E_+ 
                                  \oplus \delta m_{\nt_1} \ .
\end{eqnarray}
A very simple fit of a straight line to the upper part of the spectra 
yields a maximal energy of
$E_+ = 12.94 \pm 0.52~\GeV$ which translates to  
$ \delta m_{\stau_1} = 0.22~\GeV$ for a mass difference of
$\Delta m = 5~\GeV$.
For the scenario with $\Delta m = 3~\GeV$ one obtains 
$E_+ =  7.60 \pm 0.72~\GeV$ or $\delta m_{\stau_1} = 0.28~\GeV$.
These resolutions for very small stau-neutralino mass differences are
very encouraging and can be further improved with more refined spectra
analyses. 
All results for the $\stau_1$ mass precision obtained under different
assumptions are summarised in \tab{tab_staumasses}.

\section{Case study model D'}

The post-{\sc Wmap} model D'~\cite{pwmap} 
(parameters
$m_0=101~\GeV$, $m_{1/2}=525~\GeV$, $\tan\beta=10$, $A_0=0~\GeV$,
${\rm sign }\mu<0$) 
is a typical benchmark in the coannihilation region.
The relevant masses are the light sleptons 
$m_{\smur} = m_{\ser} = 223.9~\GeV$ and $m_{\stau_1}=217.5~\GeV$
and the neutralino $m_{\nt_1} = 212.4~\GeV$.
For the analysis the collider is chosen to operate at
$\sqrt{s}=600~\GeV$ with polarised beams of $\cP_{e^-}=+0.8$ and
$\cP_{e^+}=-0.6$, which essentially suppresses all other \susy\
production processes. The luminosity is assumed to be $\cL=300\;\fbi$.
 
From the discussion of section 3.1 it is safe to conclude that the
neutralino mass can be determined to an accuracy of 
$\delta m_{\nt_1}\simeq 0.1~\GeV$, which will be assumed in the
following stau analysis.

\subsection{Analysis of {\boldmath $e^+e^-\to\stau^+_1\stau^-_1$}}

The study of 
$e^+_L e^-_R\to \stau_1^+ \stau_1^- \to\tau^+\nt_1 \,\,\tau^-\nt_1$
proceeds along the same lines as described in section 3.2.
The cross section is  $\sigma_{\stau_1\stau_1} = 50~\fb$. The event
selection is identical except a tighter cut (6) on the tau energy
of $2 < E_\tau < 15~\GeV$, due to the small
stau-neutralino mass difference $\Delta m = 5.1~\GeV$.
The resulting overall efficiency is $7.6\,\%$.
In addition to $\ee\to\ee\tau\tau$
a hadronic background from two-photon quark production
$\gamma\gamma\to q \bar q$ is included.
The major contribution comes from charm quarks where the
final states mimic a tau pair. There is still an uncertainty on
some missing graphs concerning real or `direct'
$\gamma\gamma$ interactions, which may contribute with a similar
magnitude.   
Fortunately, the total background can be kept small at
a few percent level, and an increase by a factor of $\sim1.5$ would not
change the results significantly.
In a final experiment the $\gamma\gamma$
background will be precisely measurable and should be reliably calculable.

\begin{figure}[htb]
\begin{picture}(150,80)  (0,0)
  \put(-10,-70){
    \put(0,0){\epsfig{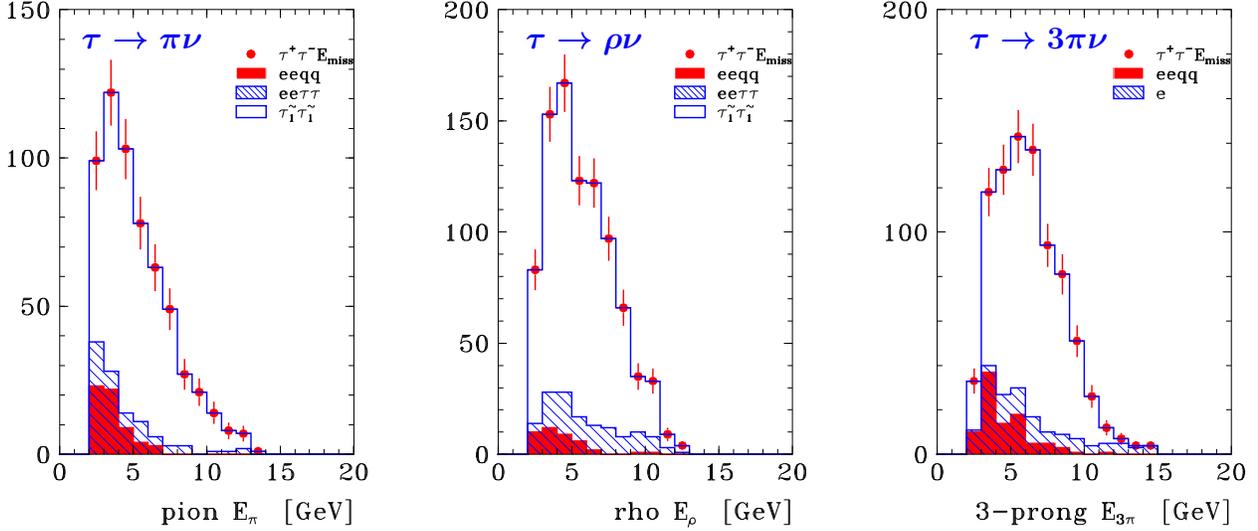} }
    \put(17,138){\color{blue} \boldmath$\tau\to\pi\nu$}
    \put(76,138){\color{blue} \boldmath$\tau\to\rho\nu$}
    \put(135,138){\color{blue}\boldmath$\tau\to 3\pi\nu$}
  }     
\end{picture}
  \caption{Hadron energy spectra $E_\pi$ of $\tau\to\pi\nu_\tau$,
    $E_\rho$ of $\tau\to\rho\nu_\tau$ and
    $E_{3\pi}$ of $\tau\to 3\pi\nu_\tau$ decays
    from the reaction
    $e^+_L e^-_R\to \stau_1  \, \stau_1   
    \to \tau^+ \nt_1 \,\, \tau^-\nt_1 $ and 
    two-photon production
    assuming head-on collision.
    Model D', $m_{\stau_1} = 217.5\;\GeV, \ \Delta m = 5.1\;\GeV$,
    $\sqrt{s}=600\;\GeV$ and $\cL=300\;\fbi$}
  \label{etaud_spectrum}
\end{figure}

The energy spectra of  $E_\pi$,  $E_\rho$ and  $E_{3\pi}$ are shown in
\fig{etaud_spectrum} for head-on collisions. One observes narrow
distributions with a small background increasing towards low energies.
From fits to the spectra assuming various $\stau_1$ mass hypothesis, 
one obtains uncertainties in the mass determination of
$\delta m_{\stau_1} = 0.19~\GeV \oplus \delta m_{\nt_1}$ 
for the $E_\pi$ spectrum, 
$\delta m_{\stau_1} = 0.12~\GeV \oplus \delta m_{\nt_1}$ 
for the $E_\rho$ spectrum and
$ \delta m_{\stau_1} = 0.14~\GeV \oplus \delta m_{\nt_1}$ for the
$E_{3\pi}$ spectrum. 
Combining the information of all measurements and folding in the
neutralino mass error ($\delta m_{\nt_1} = 0.1~\GeV$)
the stau mass can be determined as
$m_{\stau_1} = 217.5 \pm 0.15~\GeV$. 
Despite the lower event rates 
the achievable accuracy is very similar to that obtained for the
SPS~1a inspired model (see section 3.2). 
This supports the observation that narrow distributions have a somewhat
higher analysing power.
The results of the simulation of model D'
are given in \tab{tab_staumasses}.

\begin{table}[htb] \centering
\fbox{
\begin{tabular}{l c c c c}  
  scenario        & $m_{\stau_1}\ [\GeV]$ & $\delta\, m_{\stau_1}\ [\GeV]$ 
                  & $\Delta m \ [\GeV]$   & $\delta\, \Omega_{ CDM}h^2$ 
   \\[.1em]   \hdick \\[-1.em]
  SPS~1a inspired & $133.2$  & $0.14$  & $ 8 $ & $ 1.7\,\%$ \\
                  &          & $0.22$  & $ 5 $ & $ 4.1\,\%$ \\
                  &          & $0.28$  & $ 3 $ & $ 6.7\,\%$ \\[.2em]
  model D'        & $217.5$  & $0.15$  & $ 5.1 $ & $ 1.9\,\%$ \\

\end{tabular}   }
\caption{Expected accuracies on the stau mass $\delta m_{\stau_1}$
  and the resulting precision on dark matter densitiy
  $\delta\, \Omega_{ CDM}h^2$ 
  for variants of a SPS~1a inspired scenario 
  ($\cL = 200\,\fbi$ @ $400~\GeV$) and the
  benchmark model D' ($\cL = 300\,\fbi$ @ $600~\GeV$) }
\label{tab_staumasses}
\end{table}

\section{Conclusions}

Simulations of slepton production 
$\ee\to\smur\smur$ and $\ee\to \stau_1\stau_1$ in scenarios with small
slepton-neutralino mass differences $\Delta m$ are presented under realistic
experimental conditions.
Choosing the proper collider energy and beam polarisations, the decay
lepton energy spectra are analysed and allow the masses of the light
sleptons and the neutralino $\nt_1$ to be determined very precisely, 
to an accuracy of one per mil.
The results of various case studies with ranges 
$\Delta m = 8~\GeV\to 3~\GeV$
are summarised in \tab{tab_sleptons} and \tab{tab_staumasses}.

Methods are devised to efficiently suppress the most serious
background from $\gamma\gamma$ four-fermion production 
$\ee\to\ee f \bar f$, with $f = \mu,\ \tau, \ q$, which becomes more
important with decreasing $\Delta m$.
It is mandatory to have excellent veto capabilities for the
scattered $e^\pm$ down to very small angles as close as possible to
the beam pipe. 
Low angle tagging can be more easily achieved in head-on collisions
of the \tesla\ design. 
In the case of a crossing angle of $2\cdot 10\,\mrad$ the
$\gamma\gamma$ background is larger by a factor of $1.5-2$.
However, one does not expect significant degradations of the mass 
resolutions as long as the background can be kept small, 
say $\lesssim \cO( 10\%)$.

For the model D', which is typical for the coannihilation region,
the achievable $\stau$ mass resolution from energy spectra 
is $\delta m_{\stau_1} = 0.15~\GeV$.
The present analysis can be compared with an alternative method, based
on a cross section measurement of $\stau_1\stau_1$ production close to
threshold~\cite{bambade}. 
Under favourable conditions and assuming a luminosity of $500~\fbi$ an
accuracy of $0.54~\GeV$ is quoted, which is considerably worse than
the value obtained from energy spectra of the decay particles.

It is interesting to put the results in the context of a dark matter
scenario. Within the coannihilation region the dark matter content of
the universe is essentially determined by 
$\Delta m$, respectively the masses $m_{\stau_1}$ of the stau and 
$m_{\nt_1}$ of the neutralino. 
A calculation of the relic dark matter density with the program
{\sc micrOMEGAs}~\cite{micromega} and
using the uncertainties of the present study shows 
that $\Omega_{CDM}h^2$ can be predicted at the few percent level,
see \tab{tab_staumasses}.
This emphasises the important role of a future TeV Linear Collider to
provide high precision measurements.

\paragraph{Acknowledgement}
I want to thank Z.~Zhang for valuable discussions concerning some
aspects on tau detection and two-photon physics,
K.~B\"u\ss er for supplying the calculations of background caused by
beamstrahlung, and V.~Drugakov for providing the efficiency functions of
the calorimeter surrounding the beam pipe.

\newpage
%
%

\end{document}